\def\gr{$\gamma$-ray }
\def\grs{$\gamma$-rays }

\documentclass{aa}
\begin{document}
\thesaurus{08(02.02.1; 02.18.5; 09.03.2; 13.07.3)}

\title{Limits for an inverse bremsstrahlung origin of the diffuse Galactic
soft gamma-ray emission}
\author{Martin Pohl}
\institute{Danish Space Research Institute, Juliane Maries Vej 30, 2100 Copenhagen \O, Denmark}
\date{Received ; accepted }
\offprints{mkp@dsri.dk}
\titlerunning{Limits for Galactic inverse bremsstrahlung}
\maketitle

\begin{abstract}
RXTE, GINGA, and OSSE observations have revealed an intense low-energy
\gr continuum emission from the Galactic plane, which is commonly
interpreted as evidence for the possible existence of a strong flux of 
low-energy cosmic ray electrons.
In this paper I discuss the scenario of
a hadronic origin of the 
soft Galactic \gr continuum through inverse bremsstrahlung. 

A flux of
low-energy cosmic rays strong enough to produce the observed spectrum of
\grs implies substantial \gr emission at a few MeV through nuclear 
de-excitation. It is shown that
the existing limits on excess 3-7 MeV emission from the Galactic plane,
in concert with the constraints from $\pi^0$-decay \gr emission at higher energies, are in serious conflict with an inverse bremsstrahlung origin
of the Galactic soft \gr emission for any physically plausible low-energy
cosmic ray spectrum.
While in case of energetic heavy nuclei
the limits are violated by about an order of magnitude, for
a large population of low-energy protons the implied
\gr line flux and $\pi^0$-decay continuum intensity are larger than the
existing limits by at least a factor of 2.


\end{abstract}

\section{Introduction}

The Galactic plane is an extended source of $\gamma$-radiation. This has been
shown at energies $>$ 50 MeV with SAS~2 (Hartman et al. 1979), COS-B
(Strong et al. 1988), and most recently with EGRET (Hunter et al. 1997).
Observations made with COMPTEL have demonstrated that this emission extends
down to energies near 1 MeV (Strong et al. 1996). The diffuse Galactic \gr continuum at energies below 1 MeV is less well 
determined which is primarily due to the presence of a number of
point sources, many of which are variable. In an analysis of
Galactic plane observations made with OSSE (Purcell et al. 1996), it was found
that, when the contribution from the prominent point sources monitored
during simultaneous observations with SIGMA is subtracted from the 
Galactic center spectrum measured with OSSE, the diffuse emission is 
essentially identical to that measured at $l=25^\circ$. The residual
intensity is roughly
constant over the central radian of the Galaxy, but is lower by a factor 4
at $l\approx 95^\circ$ (Skibo et al. 1997). If the residual
emission is due to discrete sources, 10 sources of flux $5\cdot 10^{-3}$
ph./cm$^2$/sec/MeV at 100 keV must be present in the field of view of OSSE
to make up the spectrum. Because no such class of sources with a uniform space density in longitude is known, the emission at about 100 keV is probably of
diffuse origin. Estimates based on the luminosities
and number-flux distributions of galactic sources indicate that the point
source contribution to the hard X-ray emission from the Galactic plane is
less than 20\% (Yamasaki et al. 1997; Kaneda 1997).
Similar results have
been found in an analysis of GRIS data (Gehrels et al. 1991).
The residual source-subtracted spectrum of this emission changes from a
photon index $\alpha =1.7$ at energies above 200 keV (Strong et al. 1994)
to a photon index $\alpha =2.7$ at lower energies (Purcell et al. 1996).
Thus the soft \gr continuum from the Galactic plane is more intense than 
the extrapolation of the higher energy emission. Observations of the
Galactic ridge in the hard X-ray range with GINGA (Yamasaki et al. 1997) and
RXTE (Valinia \& Marshall 1998) indicate that the soft spectrum below 200 keV
extends down to about 10 keV energy, though the best spectral fit between 15 keV
and 150 keV gives a photon index of $\alpha =2.3$. The spectrum may also be 
represented by an exponentially absorped power-law of index $\sim 1.7$ and cut-off energy $\sim 130$ keV (Kinzer et al. 1998).

In previous papers electron bremsstrahlung has been considered the most
likely source of the soft \gr continuum.
The power in cosmic ray electrons required to produce a given amount of 
bremsstrahlung is a fixed quantity that depends only on the energy spectrum
of the radiating electrons and weakly on the ionization state of the interstellar medium. Attributing this power input to injection in cosmic ray 
electron sources it has been estimated that a power of $10^{42}-10^{43}
\,{\rm ergs\ sec^{-1}}$ in low-energy electrons is required, to maintain them
against severe Coulomb and ionization losses (Skibo et al. 1996).
This electron power exceeds the power supplied to cosmic ray nuclei by an order
of magnitude. The energy losses of the required large population of
low-energy electrons would be more than adequate to account
for the observed hydrogen ionization rate in the interstellar medium 
(Valinia \& Marshall 1998). Understanding the Galactic continuum emission
below 1 MeV is therefore of utmost importance to pin down the most relevant 
particle acceleration process and the ecosystem interstellar medium.

Recently, inverse bremsstrahlung (Boldt \& Serlemitsos 1969)
has been proposed as an alternative
to electron bremsstrahlung as basic radiation process for the low-energy
Galactic \gr continuum (Valinia \& Marshall 1998). The source power required 
to maintain the radiating nucleons against the energy losses is similar
to that in the case of electrons, but two aspects seem to be in
favor of this scenario. First, the acceleration process does not need to favor electrons, thus shock acceleration or gyroresonant interactions with
incompressive shear Alfv\'en waves could be invoked, similar to standard 
scenarios for solar flares (for a review see Ramaty \& Murphy 1987).
Second, the range of
50 MeV nucleons is about 1 g/cm$^2$ as opposed to $10^{-3}$ g/cm$^2$ for
25 keV electrons (Berger \& Selzer 1964), which allows nucleons to
diffuse away from their sources and to cause
radiation with the observed latitude distribution with 5$^\circ$ FWHM.

In this paper I discuss the scenario of
a hadronic origin of the 
soft Galactic \gr continuum in more detail. In the
next section I calculate the emissivity for inverse bremsstrahlung
and the nuclear excitation rate of low-energy cosmic ray nucleons.
In the third section it is shown that even for very hybrid nucleon spectra
the observed limits for nuclear \gr line emission and $\pi^0$-decay continuum
emission are in conflict with an inverse bremsstrahlung origin of the
Galactic low-energy \gr continuum.

\section{Emission from low-energy nucleons}

\subsection{Inverse bremsstrahlung}

The combined data of OSSE and RXTE of X-ray/\gr emission
between 15 keV and 150 keV
from the inner radian of the Galactic plane can be described by 
a differential photon spectrum  
(Valinia \& Marshall 1998)
\begin{equation}
I(\epsilon )=(1.8\cdot 10^{-3})\,\epsilon^{-2.3}
\end{equation}
where the photon energy $\epsilon$ is in units of $m_e c^2$ and all other
units in cgs.

Let $N(E) = N_0\,E^{-s}$ be the differential spectrum of radiating nucleons
over the range $\left[E_1,E_2\right]$,
where $E$ is in units of the particle's rest mass $A\, m_p\,c^2$.
We may use the non-relativistic limit of the Bethe-Heitler cross section
to write the cross section for inverse bremsstrahlung
(Boldt \& Serlemitsos 1969)
\begin{equation}
\epsilon \left({{d\sigma}\over {d\epsilon}}\right)=
{{2\,\alpha\,Z^2\,\sigma_T}\over {\pi\,E}}\ \ln\left(
\sqrt{{E}\over {\epsilon}}+\sqrt{{{E}\over {\epsilon}}-1}\right)
\end{equation}
$$
{\rm where}\quad E\ge \epsilon 
$$
The emissivity of inverse bremsstrahlung therefore is
\begin{eqnarray}
{{dq}\over {d\epsilon}}&=&
c\,n_e\,N_0 \int_{{\rm max}(\epsilon,E_1)} dE\ 
\beta\,E^{-s}\,{{d\sigma}\over {d\epsilon}}\nonumber \\
&=& (10^{-16})\, n_H\,Z^2\,N_0\,\epsilon^{-2.3} 
\end{eqnarray}
$$ {\rm for} \quad \ E_1 \le \epsilon \quad {\rm and}\ n_e =\sum_{Z=1}^{26}
n_Z\,Z\ \simeq 1.2\, n_H 
$$
and the spectral index of radiating particles $s=1.8$. By comparison with
the observed flux (Eq.1) we infer
\begin{equation}
N(E) = (1.8\cdot 10^{13}) \,(C\, n_e)^{-1}\,Z^{-2}\ E^{-1.8} \ ,\ \ E\ge 0.03
\end{equation}
where the constant $C$ describes the geometry. It is basically the ratio 
of flux from the inner radian and emissivity. Eq.4 displays the differential number spectrum of a nucleon of charge number $Z$, that is required to make up
the observed soft \gr continuum (Eq.1) through inverse brems\-strah\-lung.
Secondary electron bremsstrahlung contributes at a level of $\la$10\%
to inverse bremsstrahlung for target material with solar
abundance and is thus negligible (Abraham et al. 1966,
Ramaty et al. 1997).
In the next section I will calculate the flux of nuclear line emission 
produced by such a population of low-energy cosmic ray nuclei.

\subsection{Nuclear excitation}
The strongest contribution to Galactic \gr line emission can be expected from 
de-excitations of carbon  nuclei at 4.4 MeV and oxygen nuclei
at 6.1 MeV. The cross sections for nuclear excitation above 10 MeV/nuc.
are roughly
\begin{equation}
\sigma(C)\simeq 3.1\cdot 10^{-25} \left({{E}\over {\rm 0.01}}\right)^{-1.2}\ 
\ {\rm cm^2} 
\end{equation}
\begin{equation}
\sigma(O)\simeq 1.6\cdot 10^{-25} \left({{E}\over {\rm 0.01}}\right)^{-0.5}\ 
\ {\rm cm^2} 
\end{equation}
where $E$ is again in units of $A\, m_p\,c^2$ (Ramaty et al. 1979). The flux of
\gr line emission from the inner radian of the Galactic plane then is
\begin{equation}
I = \eta \,n_H\, c\,C\int dE \beta\,N(E)\,\sigma
\end{equation}
where $\eta$ is the abundance of the target material relative to that of
hydrogen. 

If the radiating particles are heavy nuclei, then the efficiency of
inverse bremsstrahlung would be increased by the factor $Z^2$ but
the target material hydrogen would have an abundance $\eta=1$, so that
for a given flux of inverse bremsstrahlung energetic protons would produce 
roughly an order of magnitude less flux in \gr lines than do energetic
Carbon or Oxygen nuclei.
If only energetic carbon nuclei were responsible for the entire observed
soft \gr continuum then the implied flux of \gr line emission from the inner radian of the Galactic plane would be
\begin{equation}
I_C\simeq 3.4\cdot 10^{-3}\ \ {\rm ph./cm^2/sec}
\end{equation}
Likewise we obtain for energetic oxygen nuclei
\begin{equation}
I_O\simeq 4\cdot 10^{-3}\ \ {\rm ph./cm^2/sec}
\end{equation}
Due to the smaller excitation cross sections the implied
\gr line fluxes for other heavy nuclei would be lower, so that
depending on the abundances of heavy nuclei in low-energy cosmic rays
in the inner Galaxy an actual "average" \gr line flux would be
$\ga 10^{-3}\ {\rm ph./cm^2/sec}$.

If on the other hand a large population of low-energy protons is responsible
for the soft \gr emission, then the implied flux of \gr line emission from the inner radian of the Galactic plane would be
\begin{equation}
I_H\simeq 2.6\cdot 10^{-4}\ \ {\rm ph./cm^2/sec}
\end{equation}
assuming a solar abundance of Carbon ($\eta_C =3.6\cdot 10^{-4}$) and
Oxygen ($\eta_O =8.5\cdot 10^{-4}$) (Grevesse \& Anders 1989).
Note that these estimates for the \gr line flux have been calculated
using the limits
of the particle spectrum as in Eq.4, i.e. assuming
that no energetic particles exist with energies below $E_1=0.03$,
and therefore they should be taken as lower limits.

If the low-energy protons are enriched by heavier nuclei with a
differential intensity ratio corresponding to solar abundances, the estimate
in Eq.10 would be doubled and one would observe broad and narrow lines at
similar flux levels. 

\section{Discussion}

The results of the OSSE and COMPTEL experiments on \gr line
emission following the de-excitation of carbon and oxygen nuclei are still
preliminary.
The OSSE team has reported a 3$\sigma$ upper limit for narrow lines from
the inner radian of the Galactic plane (Harris et al. 1996)
\begin{equation}
I_{C,O}({\rm OSSE}) \le 1.44\cdot 10^{-4}\ \ {\rm ph./cm^2/sec}
\end{equation}
For broad lines, i.e. energetic Carbon and Oxygen nuclei,
the upper limit is about 50\% higher.
COMPTEL data have revealed some indication for an excess 3-7 MeV flux
from the Galactic ridge at a level of (Bloemen \& Bykov 1997, Bloemen et al. 1997)
\begin{equation}
I_{C,O}({\rm COMPTEL}) \la 10^{-4}\ \ {\rm ph./cm^2/sec}
\end{equation}
These limits strongly exclude a large population of low-energy heavy nuclei
causing the observed soft \gr continuum. Inverse bremsstrahlung of protons 
seems to be weakly excluded as the implied \gr line flux is a factor of 2
higher than the 3$\sigma$ OSSE limit. 

The main sources of uncertainty in the implied \gr line flux are:
a) the abundance of Carbon and Oxygen in the inner Galaxy, b) the fraction of
the soft \gr continuum which is truly diffuse emission, and c) the true
low-energy limit of the power-law spectrum in Eq.1.

The metallicity in the inner Galaxy is higher than in solar vicinity
(Shaver et al. 1983), so that we can expect Carbon and Oxygen abundances
to be higher than solar. This would lead to a higher
\gr line flux and thereby increase the discrepancy between implied flux
and the observational limits. The broad longitude distribution of the soft
\gr continuum indicates that it is emitted roughly homogeneously throughout
the Galactic disk. In the remaining discussion
we may therefore assume a metallicity like that near the molecular
ring at about 4 kpc galactocentric radius, which is roughly a factor
2 higher than solar (Shaver et al. 1983).

The main contribution to the implied \gr line flux in Eq.10 comes from 
excitations of ambient Oxygen nuclei. Due to the weak energy dependence of
the Oxygen excitation cross section the \gr line flux depends on the
lower limit of the proton spectrum roughly as $\propto E_1^{-0.8}$.
Therefore, the predicted \gr line flux would harmonize with
the observational limits only if $E_1$ were $\ge 0.1$, which means
that the continuum emission below 50 keV could not be entirely
caused by inverse bremsstrahlung.

For energies above E=0.3 the \gr continuum emission following 
pion production and decay provides another constraint on the number of 
energetic protons in the Galaxy. The combined data of EGRET, OSSE, and
COMPTEL show that the \gr luminosity of the Galaxy at a few hundred MeV
is about twice as much as that at 30 keV (Purcell et al. 1996). For
protons with less than 100 MeV energy we can assume the Galaxy to act as
a thick target (Pohl 1993), thus the particle spectrum is only determined by the
energy losses. From the well-known energy dependence of Ionization and
Coulomb losses (Heitler 1954; Butler \& Birmingham 1962) we can infer
that an $Q\propto E^{-3.3}$ injection spectrum is required to maintain an
$N\propto E^{-1.8}$ equilibrium spectrum. The luminosity at a few hundred MeV 
photon energy then should scale to that at 30 keV like the ratio of
the radiation efficiency times injected energy in the appropriate energy bands.
Given the results for the mean density and escape lifetime of cosmic rays
(Webber et al. 1992) we can estimate that for $\pi^0$-production
$\eta \simeq 0.02$. For inverse bremsstrahlung at 30 keV
the radiation
efficiency can be calculated to be $\eta \simeq 4\cdot 10^{-5}$.
Therefore we obtain
\begin{equation}
{{L_{\pi}}\over {L_{IB}}} = {{(\eta\,Q\,E^2)_{\pi}}\over 
{(\eta\,Q\,E^2)_{IB}}} \, \simeq\, 20
\end{equation}
which exceeds the observed ratio of two by an order of magnitude.
This implies that the proton spectrum 
(Eq.4) would have to cut off sharply at energies near the pion production
threshold of about $E\simeq 0.3$, corresponding to a cut-off in the inverse
bremsstrahlung spectrum near 100 keV.

\subsection{Exponentially absorbed particle spectra}
We have seen that the limits for \gr line emission constrain the particle
spectrum at low energies whereas \gr continuum emission subsequent to
pion production constrains at high energies. In this subsection
I will investigate for two extreme cases if exponentially absorbed
particle spectra can be found, which harmonize with both constraints
\emph{and} reproduce the observed soft \gr continuum via inverse
bremsstrahlung emission. The abundances of Carbon and 
Oxygen will be assumed to be two times solar.

Suppose a proton spectrum of the form
\begin{equation}
N(E)=N_0\, E^{-1}\,\exp(-\,3.3\,E)
\end{equation}
where the argument of the exponential has been chosen such as to reproduce
the observed soft \gr continuum. The choice of spectral form is 
entirely hybrid. If low-energy protons with a spectrum as in Eq.14 were
responsible for the observed soft \gr continuum via inverse bremsstrahlung
then the implied $\pi^0$-decay \gr luminosity at a few hundred MeV would be
2.5 times that of the 30 keV inverse bremsstrahlung emission, roughly in 
accordance
with the observed ratio of two. However, the narrow \gr line emission would 
have a flux of $10^{-3} \ {\rm ph./cm^2/sec}$, clearly above the
observational limits. Thus the spectrum Eq.14 would be too soft to harmonize with the low-energy constraints.

The flux of low-energy (E=0.01) protons may be reduced if distinct sources
provide an injection spectrum like Eq.14 such that energy losses within the
sources can be neglected. Then the overall particle spectrum would result
from a balance equation for energy losses and injection (Eq.14) as
\begin{equation}
N(E)=N_0\, \sqrt{E}\int_E du\ u^{-1}\,\exp(-{{u}\over {0.25}})
\end{equation}
where the argument of the exponential has been readjusted to reproduce
the observed low-energy \gr spectrum. The implied narrow \gr line flux would be
about $2\cdot 10^{-4} \ {\rm ph./cm^2/sec}$,
but the implied $\pi^0$-decay \gr luminosity at a few
hundred MeV is 30 times that of the 30 keV continuum, clearly
in conflict with the data.

An exponential cut-off in the low-energy proton spectrum seems to be 
insufficient to satisfy the constraints at low and high energy. Therefore only
a quasi-monoenergetic injection of pure protons would harmonize with the data.
I do not regard this a physically plausible scenario.

\section{Conclusions}
 
An inverse bremsstrahlung origin of the intense soft Galactic
\gr continuum emission implies substantial \gr line emission through 
nuclear de-excitation and it is thus testable.

The existing limits on Galactic \gr line emission exclude a large population
of low-energy heavy nuclei.

For low-energy protons the current
observational limits are in conflict with a large population of Galactic
cosmic ray proton both below 40 MeV and above 400 MeV. 
It seems that no physically plausible proton spectrum can be
constructed which would reproduce the observed low-energy \gr continuum via
inverse brems\-strah\-lung emission and harmonize with the existing limits for 
nuclear \gr line emission and $\pi^0$-decay \gr emission.

\end{document}